\documentclass[aps,
final,%
notitlepage,%
onecolumn,%
centertags,
floatfix, pra]%
{revtex4-1}
\usepackage{showkeys}
\usepackage{graphicx}
\usepackage{mathtools}
\usepackage[normalem]{ulem}
\usepackage{cancel}
\usepackage{hyperref}
\usepackage{amsmath,amsthm,amssymb,amscd}
\usepackage{tikz}
\usepackage{graphicx,graphbox}
\usepackage{multirow}
\usepackage{ulem}
\usepackage{dsfont}

\DeclareMathOperator{\Tr}{Tr}
\usepackage{color}

\newcommand{\rd}{{\mathrm d}}

\newcommand{\C}{\mathbb C}
\newcommand{\rdt}{\mathrm{d}^2}

\newcommand{\e}{{\mathrm e}}
\newcommand{\Ccal}{{\mathcal C}}

\newcommand{\Pcal}{{\mathcal P}}

\newcommand{\rhoG}{\rho_{\textrm G}}
\newcommand{\OS}{{S_{\mathrm{o}}}}

\newcommand{\vertiii}[1]{{\left\vert\kern-0.25ex\left\vert\kern-0.25ex\left\vert #1 
    \right\vert\kern-0.25ex\right\vert\kern-0.25ex\right\vert}}
\newcommand{\tR}{t_{\textrm{R}}}

\newcommand{\eq}[1]{\begin{equation} #1 \end{equation}}
\newcommand{\tr}{\mathrm{Tr}}
\newcommand{\nbarinf}{\overline n_\infty}
\newcommand{\eqarray}[1]{\begin{eqnarray} #1 \end{eqnarray}}
\newcommand{\ket}[1]{\vert #1 \rangle}
\newcommand{\bra} [1] {\langle #1 \vert}

\newcommand{\ketbra}[2]{| #1 \rangle \langle #2 |}

\newcommand{\mean}[1]{\langle #1 \rangle}

\newtheoremstyle{break}
{\topsep}{\topsep}%
{\itshape}{}%
{\bfseries}{}%
{\newline}{}%
\theoremstyle{break}

\newcommand{\xthetastar}{x_{\theta_*}}
\newcommand{\pthetastar}{p_{\theta_*}}	
\newcommand{\QCStwo}{{\mathcal C}^2}
\newcommand{\QCStwoG}{{\mathcal C}^2_{\mathrm G}}

\makeatletter
\newcommand{\oset}[3][0ex]{%
	\mathrel{\mathop{#3}\limits^{
			\vbox to#1{\kern-11\ex@
				\hbox{$\scriptstyle#2$}\vss}}}}
\makeatother

\begin{document}

\title{
Quadrature coherence scale driven fast decoherence of bosonic quantum field states
}

\author{Anaelle Hertz$^{1}$, Stephan De Bi\`evre$^{2}$}

\address{
$^1$Univ. Lille, CNRS, UMR 8523, PhLAM - Physique des Lasers Atomes et Mol\'ecules, F-59000 Lille, France\\
$^2$Univ. Lille, CNRS, UMR 8524, Inria - Laboratoire Paul Painlev\'e, F-59000 Lille, France
}


\begin{abstract}
 We introduce, for each state of a bosonic quantum field, its quadrature coherence scale (QCS), a measure of the range 
of its quadrature coherences. Under coupling to a thermal bath, the purity and QCS are shown to decrease on a time scale inversely proportional to the QCS squared. 
The states most fragile to decoherence are therefore those with quadrature coherences far from the diagonal. 
We further show a large QCS is difficult to measure since it induces small scale variations in the state's Wigner function.  These two observations imply a large QCS constitutes a mark of ``macroscopic coherence''. 
Finally, we link the QCS to optical classicality:  optical classical states have a small QCS and a large QCS implies strong optical nonclassicality. 


\end{abstract}
\pacs{vvv}

\maketitle
\section{Introduction}
Both  in order to obtain an ever better understanding of quantum physics and to solve problems in quantum information theory, there is continued interest in the exploration of the classical-quantum boundary and the identification of those quantum states exhibiting specifically quantum features,
such as coherence and interference that 
cannot be explained with classical mechanics and/or classical probability theory. It has been shown on example states in model systems~\cite{Joos1985, Zurek93, dodonov00, SpohnDurr,Zurek2003,  hara13} that fast decoherence results from the interaction of the system with its environment when the system is suitably ``macroscopic''. It is therefore
much harder to generate, maintain and detect coherence on a macroscopic scale than on a microscopic one. 
These results contribute to clarifying why the observation of coherent superpositions is not part of our every day experience and why building large scale quantum computers is a major challenge.

To render the previous observations quantitative and general, several different characterizations have been proposed of the ``coherence''~\cite{bacrpl14, streltsov17}, ``large-scale quantum coherence''~\cite{frsagi15, yave16, fryagi18}, ``macroscopic coherence''~\cite{yave16, kwpataahje18}, ``quantum macroscopicity''~\cite{le02, frdu12, yave16}, ``macroscopic quantumness''~\cite{leje11, go11, leje11b, ousefrgisa15, yave16} and ``macroscopic distinctness''~\cite{le02, ousefrgisa15, segisa14} of quantum states. Resource theories for those closely related properties of states have also been developed~\cite{bacrpl14, gi14, yave16, streltsov17, kwpataahje18}.   
An important and to the best of our knowledge 
unaddressed question concerning 
these quantities 
is to evaluate the rate at which they decrease when the system is coupled to its environment: i.e. to evaluate 
their sensitivity to environmental decoherence.

We will address the latter question for the states of a bosonic quantum field for which we introduce the quadrature coherence scale (QCS), defined as a measure of the scale on which the coherences of its quadratures are appreciable (see~\eqref{eq:QCS2}). As we will show, a small QCS means the coherences \emph{for all} quadratures are small far from the diagonal. We call such states \emph{quadrature quasi-incoherent}. A large QCS means on the contrary that, given any pair of conjugate quadratures, at least one has appreciable coherences far from the diagonal. In addition, it implies the state is strongly optically nonclassical.

We show the QCS of a state, if initially large, decreases fast when the system is coupled to an environment. The corresponding  characteristic time scale is inversely proportional to the (square of the) QCS itself. Purity loss takes place on a similar time scale. Therefore,  the states very sensitive to environmental decoherence are precisely those with a large QCS. 
 This result generalizes known results on the decoherence of optical cat states~\cite{Zurek93, dodonov00, SpohnDurr, Zurek2003, hara13} to all mixed or pure states of the field mode. 
 We further establish that states with a large QCS are hard to observe since their Wigner functions have small scale structures.
The QCS thus furnishes a physical parameter that measures  the ``coherence size'' of the state and that is directly related to the decoherence rate.

\section{Coherence,  interference and coherence scale}  To reveal the quantum nature of a state $\rho$, one may proceed as follows. Consider two noncommuting observables $A$ and $B$ that we suppose for simplicity to have associated orthonormal eigenbases $|a_i\rangle$, $|b_m\rangle$. Let 
$p_A(a_i)=\langle a_i|\rho|a_i\rangle$ and $p_B(b_m)=\langle b_m|\rho|b_m\rangle$. Then 
\begin{equation}\label{eq:interference}
p_B(b_m)=p_B^{\textrm{diag}}(b_m) +\sum_{i\not=j}\langle b_m|a_i\rangle\langle a_j|b_m\rangle\langle a_i|\rho|a_j\rangle.
\end{equation}

\vspace{-5pt}\noindent Here $p_B^{\textrm{diag}}(b_m) =\sum_i |\langle b_m|a_i\rangle|^2 p_A(a_i)$
 is of  ``classical'' nature in the sense that it is the term expected from an application of 
 classical probability theory. The second term -- the interference term -- is of typical quantum nature. It is absent when $\rho$ has no off-diagonal matrix elements $\langle a_i|\rho|a_j\rangle$, the so-called ``coherences.'' For an overview of measures and monotones of coherence for observables with discrete spectrum, we refer to~\cite{streltsov17}, and references therein. If $\rho$ is diagonal in the $A$ basis ($\rho$ is then said to be $A$-incoherent) or if its coherences in this basis are small, the quantum nature of $\rho$ is not revealed in this manner and one can then say that, in this restricted sense, the state ``behaves classically''.  The state's quantum nature may however still be revealed by another choice of observables. In this view, there is no such thing as ``the'' nonclassical nature of a state, but rather the degree to which various measurements can reveal its quantum nature, notably through interference terms. The relation of this analysis to the independent notion of optical (non)classicality of the states of a bosonic quantum field mode will be discussed below.

As will become clear below, for our purposes, it is not so much the size of the coherences as their location which is important. To evaluate how far from the diagonal the coherences occur, we write $\Pcal=\Tr \rho^2$ for the purity, and consider the probability density on the $(a,a')$-plane:
$$
\mu(a,a')=\sum_{i,j} \frac{|\langle a_i|\rho|a_j\rangle|^2}{\Pcal}\delta(a-a_i)\delta(a'-a_j). 
$$
It describes the spatial repartition of the matrix elements of $\rho$ and in particular of its coherences.  We define the $A$-coherence scale $\Ccal_A(\rho)$ of $\rho$ via
\begin{equation*}
\Ccal_A^2(\rho)=\hspace{-2pt}\sum_{i,j} (a_i-a_j)^2\frac{|\langle a_i|\rho|a_j\rangle|^2}{\Pcal}
=\hspace{-4pt}\int \hspace{-2pt}(a-a')^2\mu(a,a')\rd a\rd a',
\end{equation*}
which is the variance of the eigenvalue spacings of $A$ so that, when $\Ccal_A(\rho)$ is large, there are coherences far from the diagonal: $\Ccal_A(\rho)$ determines the scale on which the coherences of $\rho$ live. 
It is easy to check that for pure states  $\Ccal_A^2(\rho)=2(\Delta A)^2$. 
A simple calculation shows furthermore
$
\Ccal_A^2(\rho)=\Pcal^{-1}\Tr[\rho, A][A,\rho],
$
an expression valid also when $A$ has continuous spectrum. We stress that $\Ccal_A$ is \emph{not} a measure of the $A$-coherence of the state; it does not establish ``how much'' coherence there is, but ``where'' it is. For example, two states $|a_i\rangle+|a_j\rangle$ and $|a_i\rangle+|a_k\rangle$ have off-diagonal matrix elements of the same size, and in this sense the same ``amount'' of coherence, but their $A$-coherence scale is proportional to $|a_i-a_j|$, respectively $|a_i-a_k|$, and can therefore strongly differ~\cite{yave16}.

\section{Quadrature coherence scale-Quadrature quasi-coherence} 
We consider a state $\rho$ of a single-mode field, characterized by an annihilation-creation operator pair $a, a^\dagger$. We are interested in the coherence scale of its quadratures and define, in analogy with what precedes, its quadrature coherence scale (QCS) $\Ccal(\rho)$ through 
\begin{equation}\label{eq:QCS}
\Ccal^2(\rho)=\frac1{2\Pcal}\left(\Tr[\rho, X][X,\rho]+\Tr[\rho, P][P,\rho]\right), 
\end{equation}
where $X=\frac{a^\dagger+a}{\sqrt2},\, P=\frac{i (a^\dagger-a)}{\sqrt2}$.
 With $X_\theta=\cos\theta X+\sin\theta P$, $P_\theta=-\sin\theta X+\cos\theta P$, one has also
\begin{equation}\label{eq:CStheta}
\Ccal^2(\rho)=\frac1{2\Pcal}\left(\Tr[\rho, X_\theta][X_\theta,\rho]+\Tr[\rho, P_\theta][P_\theta,\rho]\right),  
\end{equation}
so that $\Ccal^2(\rho)$ is the average coherence scale (squared) of any pair of conjugate quadratures. Equation~\eqref{eq:QCS} implies
\begin{equation}\label{eq:QCS2}
\Ccal^2(\rho)=\frac1{2\Pcal}\left(\int (x-x')^2 |\rho(x,x')|^2\rd x\rd x'\right. +\\ \left.\int (p-p')^2|\rho(p,p')|^2\rd p\rd p' \right).
\end{equation}
Here $\rho(x,x')$ (respectively $\rho(p,p')$) is the operator kernel of $\rho$ in the $X$-representation (respectively $P$-representation).
It follows from~\eqref{eq:CStheta}-\eqref{eq:QCS2} that a large $\Ccal(\rho)$ implies that for every pair $(X_\theta, P_\theta)$ of conjugate quadratures, at least one has a large coherence scale.  Conversely, a small $\Ccal(\rho)$ implies that the off-diagonal coherences of all quadratures must be small away from the diagonal.
We stress that no state $\rho$ can be $X_\theta\,$- or $P_\theta\,$-incoherent in the sense that $\rho$ cannot be diagonal in the corresponding representation (see Appendix \ref{incoherent}).
Mixed states can nevertheless have an arbitrarily small QCS, as we will see below. For pure states, \eqref{eq:QCS} implies $\Ccal^2(\rho)=(\Delta X)^2 +(\Delta P)^2$, the so-called total noise of $\rho$~\cite{sc86}. 
 It follows that, on pure states, the QCS is larger than 1;  it reaches its minimal value of $1$ only on the coherent
states $|\alpha\rangle=D(\alpha)|0\rangle$, where $|0\rangle$ is the vacuum state and $D(\alpha)=\exp(\alpha a^\dagger-\alpha^*a)$. 
For optical cat states $|\psi_\alpha\rangle\sim (|\alpha\rangle+|-\alpha\rangle)$, a simple computation (see Appendix~\ref{purestates}) yields $\Ccal_\alpha\simeq |\alpha|$ $(|\alpha|\gg1$): ``large'' cats have a large QCS in agreement with the observation that $\rho_\alpha(x,x')=\langle x|\psi_\alpha\rangle\langle \psi_\alpha|x'\rangle$ has large off-diagonal elements in the neighbourhood of $x=-x'=\pm\alpha$ if $\alpha$ is real. We will refer to states for which $\Ccal(\rho)\leq 1$ as  \emph{quadrature quasi-incoherent} states. 
In fact, we will see below that $\Ccal^2$ has the particular feature of providing  a measure of optical non-classicality. 
It follows from~\cite{gu90, Bievre19} that the right hand side of~\eqref{eq:QCS} can be expressed in terms of the Wigner function $W(\alpha)$ or the characteristic function $\chi(\xi)$~\cite{cagl69a, cagl69b} as follows:
\begin{equation}\label{eq:QCSWch}
\Ccal^2(\rho)=\frac{\||\xi|\chi\|_2^2}{\|\chi\|_2^2}=\frac14\frac{\|\nabla W\|_2^2}{\|W\|_2^2}.
\end{equation}
Here, with $\xi,\alpha\in \C$, $\|\cdot\|_2$ stands for the $L^2$-norm, meaning for example $\|W\|_2^2:=\int|W|^2(\alpha)\rd^2 \alpha$ and
\begin{equation*}
\chi(\xi)=\mathrm{Tr}\rho D(\xi),\ W(\alpha)=\frac1{\pi^{2}}\hspace{-2pt}\int\hspace{-2pt}\chi(\xi)\exp(\xi^*\alpha-\xi\alpha^*)\rdt\xi.
\end{equation*} 
The definition \eqref{eq:QCS} and expression \eqref{eq:QCSWch} carry over to multimode systems by summing over a complete set of conjugate quadratures.

For a centered Gaussian state $\rhoG$ with covariance matrix $V=\begin{psmallmatrix}
  		2\tr\rho X^2&\tr\rho(XP+PX)\\\tr\rho(XP+PX)&2\tr\rho P^2
  	\end{psmallmatrix}$,
one finds (see Appendix~\ref{gaussian}):
 \begin{equation}\label{eq:QCSG}
 \QCStwoG=\Ccal^2(\rho_{\textrm{G}})=((\Delta X)^2 +(\Delta P)^2)\Pcal^2=\frac12\Tr V^{-1}.
 \end{equation}
It follows that Gaussian mixed states can have an arbitrarily small QCS. This can happen even if the total noise 
\begin{figure}
	\hspace{10pt}\includegraphics[align=t, hsmash=c, width=0.3\columnwidth]{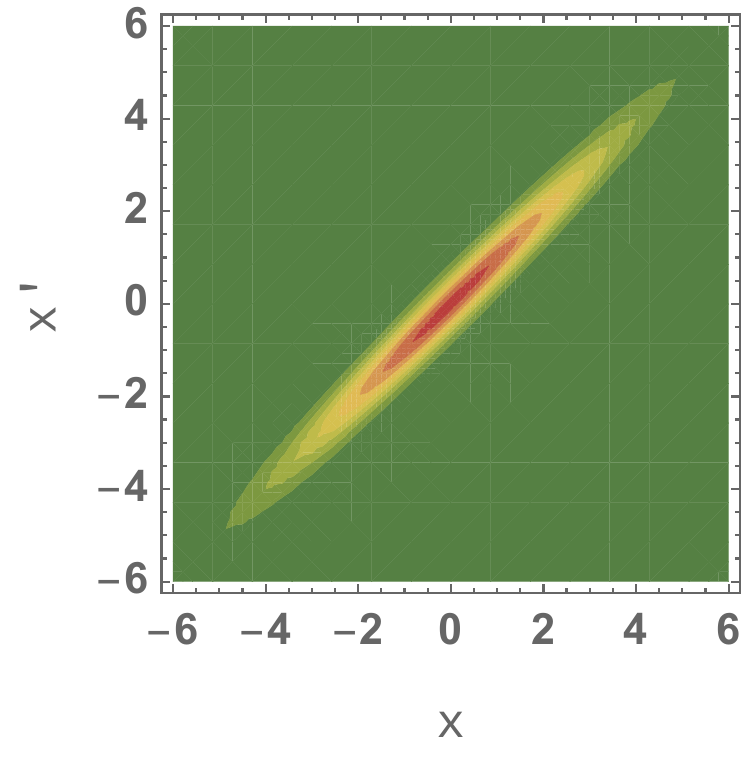}
	\hspace{5pt}
	\includegraphics[align=t, width=0.3\columnwidth]{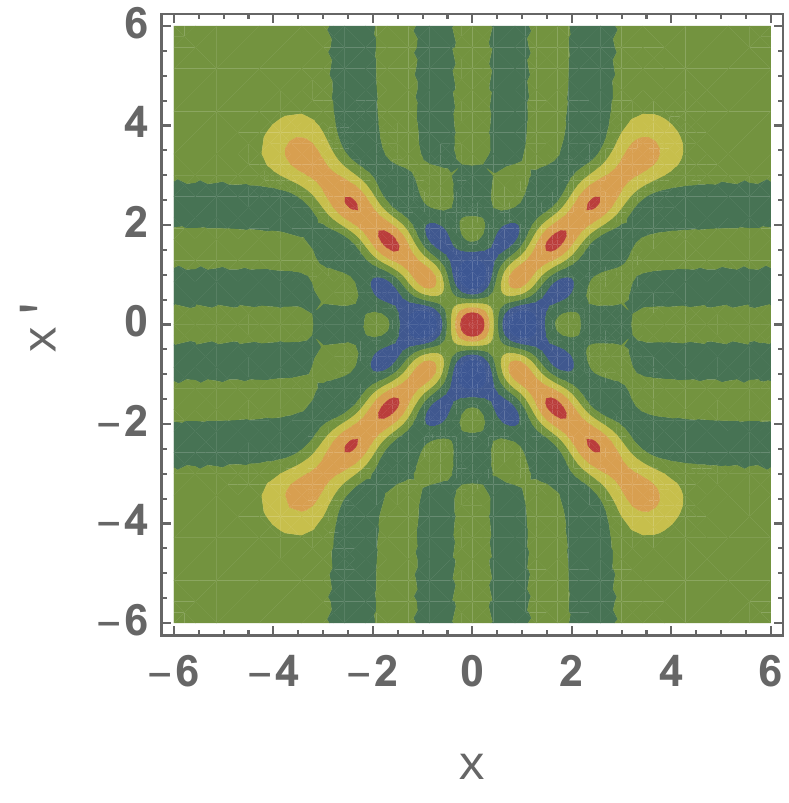}\hspace{70pt}
	\includegraphics[align=t, width=0.076\columnwidth]{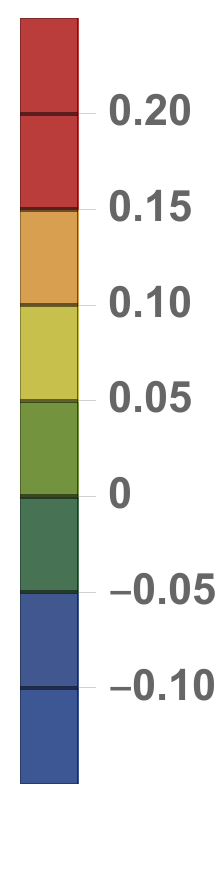}
	\vspace{-0pt}
	\caption{Plots of $\rho(x,x')$. Left panel: thermal state with $\overline n=5$. Right panel: even state $\rho_M$, with $M=4$ ,$\bar n=5$.\label{fig:QCS}}
\end{figure}
is very large. One notes for example  in Fig.~\ref{fig:QCS} that the coherences of the thermal state  with mean photon number $\overline n=5$ are concentrated along the diagonal. This reflects the fact that for thermal states $\Ccal(\rho_{\textrm{th}})=(1+2\overline n)^{-1/2}$, which follows from~\eqref{eq:QCSG}.   
We note that for Gaussian states
 $4\Ccal_{\mathrm{G}}^2$ coincides with the sum of the quantum Fisher information of two conjugate quadratures (see Appendix~\ref{gaussian}), which is known to provide a useful lower bound for proposed measures and monotones of quantum macroscopicity~\cite{yave16} and  nonclassicality~\cite{yabithnaguki18}. On non-Gaussian states, however, the two quantities can differ greatly (for an example see \cite{Bievre19}).
 
 As an example of non-Gaussian states we consider the family of  even states, with $M$ a positive integer:
\begin{equation*}\label{eq:evenM}
\rho_M=\frac{1}{M}\sum_{k=1}^M |2k\rangle\langle 2k|.
\end{equation*}
One has $\Ccal(\rho_M)=\sqrt{2M+3}$~\cite{Bievre19} and Fig.~\ref{fig:QCS} shows that, indeed, the coherences have a large off-diagonal branch that can be checked to grow as $\sqrt{2M}$, as expected. Since $\Pcal_M=M^{-1}$, this shows that very strongly mixed states can have a very large QCS. Other examples of this phenomenon are the strongly squeezed thermal states for which a very small purity can be compensated by a very large total noise (see~\eqref{eq:QCSG} and Appendix~\ref{gaussian}). 

\section{Environment induced quadrature coherence scale loss} 

\begin{figure}
	\begin{tikzpicture}
	\node[anchor=south west,inner sep=0] (image) at (-10,0) {	\includegraphics[hsmash=r, width=0.4\columnwidth]{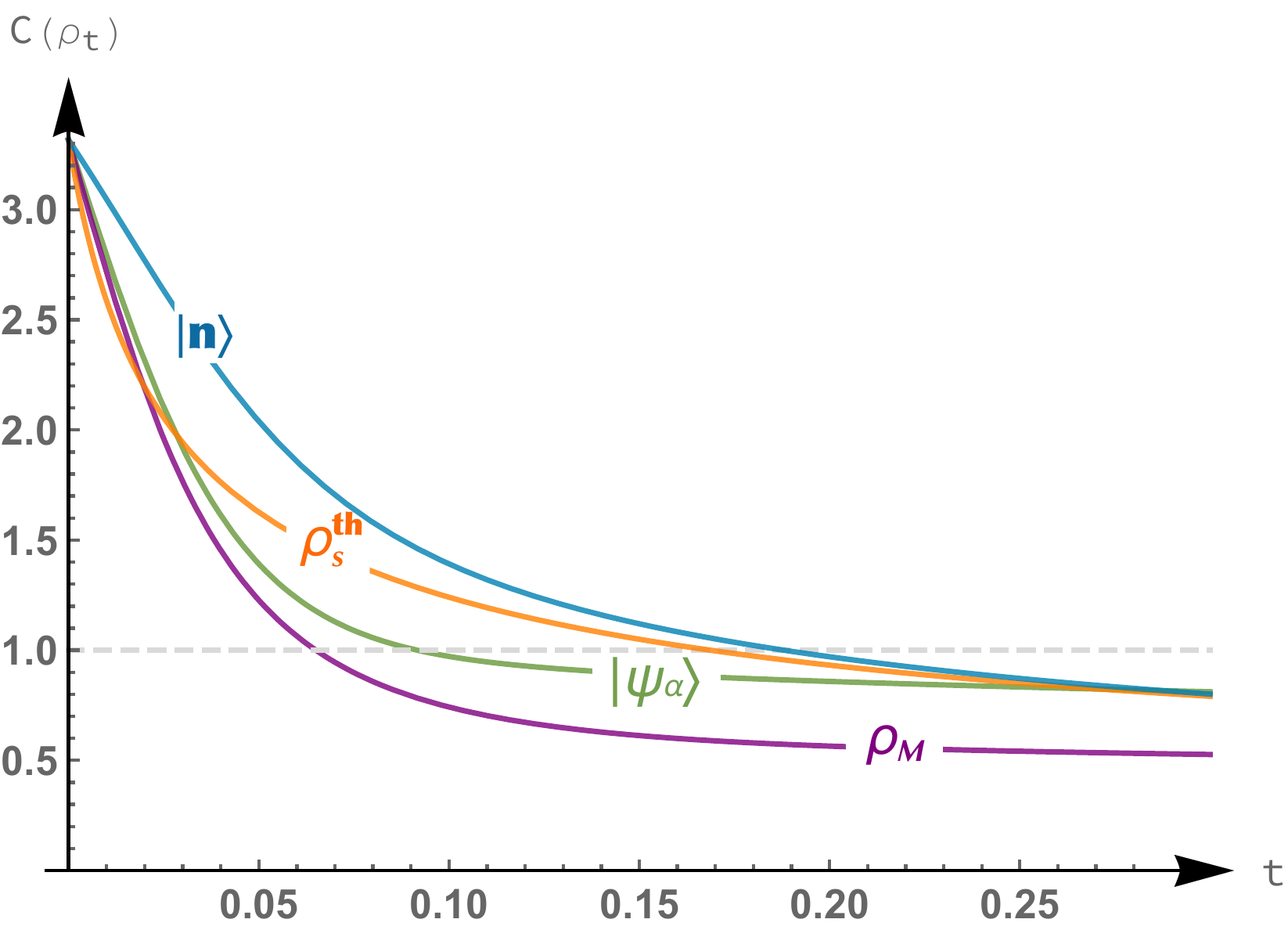}};
	\begin{scope}[x={(image.south east)},y={(image.north west)}]
	\node[anchor=south west,inner sep=0] (image) at (0.1,0.6) {\footnotesize
		\begin{tabular}{ c||c|c|c|c| } 
		& $|n\rangle $& $|\psi_\alpha\rangle $ & $\rho_M$& $\rho_s^{th}$ \\ 
		\hline\hline
		$\tau_{\Ccal}$  & 0.07 & 0.038 & 0.033 & 0.047  \\ 
		\hline
		$\tau_{\Ccal, \mathrm{appr.}}$ & 0.064 & 0.032  & 0.026 & 0.048  \\ 
		\hline
		\end{tabular}\normalsize};
	\end{scope}
	\end{tikzpicture}
	\caption{Evolution of $\Ccal(\rho_t)$  under the dynamics~\eqref{eq:QCSeqn} of an initial Fock state $|n\rangle$ ($n=5$), a squeezed thermal state ($V=1.8 \begin{psmallmatrix}
		e^{-2r}&0\\0&e^{2r}	\end{psmallmatrix},\,\, r=\cosh^{-1}(19.8)/2\approx 1.84$), an optical cat state $|\psi_\alpha\rangle\sim (|\alpha\rangle+|-\alpha\rangle)$ ($\alpha\approx 2.24$) and an even state $\rho_M$ ($M=4$). $\Ccal(0)=\sqrt{11}$ for all states shown. $\overline n_\infty = 1$. $t_R=1$. The table show the numerically exact value of the half-life $\tau_{\Ccal}$ of the QCS and its approximation obtained by Eq.~\eqref{eq:halflifeQCS} for the first three columns and Eq.~\eqref{eq:halflifeQCSGaussian} for the Gaussian state. \label{fig:focksqthQCSt}}
\end{figure}
We consider a field  weakly coupled to a thermal bath through the standard master equation in Lindblad form~\cite{scla67, Joos1985,  alicki87, dodonov00, hara13}
%
\begin{equation*}\label{eq:lindblad}
\frac{d}{dt}\rho(t)=-i\omega [a^\dag a,\rho(t)]+\frac{1}{2}\gamma\left\{[a\rho(t),a^\dag]+[a,\rho(t)a^\dag]\right\}
+\frac{1}{2}\delta\left\{[a^\dag\rho(t),a]+[a^\dag,\rho(t)a]\right\},
\end{equation*}
where $\gamma>\delta\geq0$. This dynamics converges to a thermal state with mean photon number $\bar n_\infty=\delta \tR$, where   $\tR=(\gamma-\delta)^{-1}$ is the relaxation time. 
Purity evolution is determined by $\dot \Pcal(t)= \frac1{\tR}\left[1-(2\nbarinf+1)\Ccal^2(t)\right]\Pcal(t)$. Using the affine approximation to $\Pcal(t)$ at small $t$ shows the purity half time  $$\tau_{\Pcal}\approx\frac12\frac1{(2\nbarinf+1)\Ccal^2_0-1}\tR,$$ 
provided $\Ccal_0^2=\Ccal^2(0)>1$: the purity half-life decreases as $\Ccal_0^{-2}$ when the QCS is large. This approximation gives the right order of magnitude (see Appendix~\ref{DiffEq}) and reduces to the known result for pure states~\cite{Zurek93, dodonov00, Zurek2003}.
Simultaneously with the purity loss, there is QCS loss. Indeed, the time evolution of the QCS, and in particular its sharp initial drop (Fig.~\ref{fig:focksqthQCSt}), can be explained by analyzing the differential equation for $\Ccal(t)$ (see Appendix~\ref{DiffEq}):
\begin{eqnarray}\label{eq:QCSeqn}
\dot \Ccal(t) &=& \frac1{2\tR}\left[1-\kappa(t)(2\nbarinf+1)\Ccal^2(t)\right]\Ccal(t),
\end{eqnarray}
 with $\kappa(t)=\left(\frac{\langle\langle\xi^4\rangle\rangle_t}{\langle\langle\xi^2\rangle\rangle_t^2}-1\right)$, $
\langle\langle\xi^{2k}\rangle\rangle=\int |\xi|^{2k} \frac{|\chi(\xi)|^2}{\|\chi\|_2^2} \rd x$.
Hence,  the half-life $\tau_\Ccal$ of the QCS is given approximately by (Fig.~\ref{fig:focksqthQCSt})
\begin{equation}\label{eq:halflifeQCS}
\tau_{\Ccal}\approx-\frac12\frac{\Ccal(0)}{\dot{\Ccal}(0)}= \frac{1}{\kappa_0(2\nbarinf+1)\Ccal^2_0-1}\tR.
\end{equation}
where $\kappa_0=\kappa(0)$. For Gaussian states, more precise estimates can be obtained from a more detailed computation (see Appendix~\ref{gaussian}):
\begin{eqnarray}
\tau_{\mathcal P, \mathrm{G}}&\approx&\frac{2^{\kappa_0}-1}{\left((2\overline n_\infty +1)\Ccal_0^2-2^{\kappa_0}\right)\kappa_0}\tR,\nonumber\\
\tau_{\Ccal, \mathrm{G}}&\approx& \frac{3}{\kappa_0(2\bar n_\infty+1)\mathcal{C}_0^2-4} t_{\mathrm R}\label{eq:halflifeQCSGaussian}.
\end{eqnarray}
Comparing~\eqref{eq:halflifeQCS} to~\eqref{eq:halflifeQCSGaussian},  one sees that for the same value of $\Ccal_0\gg1$ and $\kappa_0$, a Gaussian state is less sensitive to decoherence than the non-Gaussian states considered above.    A further calculation (see Appendix~\ref{gaussian}) permits to  determine the time $\tau_{1, \mathrm{G}}$ at which the state becomes quasi-incoherent, \emph{i.e.} $\mathcal{C}(\tau_{1, \mathrm{G}})=1$. It is, remarkably, to leading order in $\Ccal_0^{-2}$, independent of the QCS:
$$
\tau_{1, \mathrm{G}}\approx\left(\ln\left(\frac{\kappa_0(2\bar n_\infty+1)}{\kappa_0(2\bar n_\infty+1)-1}\right)-\frac{1}{\mathcal{C}_0^2\kappa_0(2\bar n_\infty+1)}\right)\tR.
$$
These results show in all generality that the purity loss and the destruction of the large scale quadrature coherences of any initial state are determined by the temperature of the environment and by two parameters characteristic of the initial state: the QCS $\Ccal_0$ and $\kappa_0$. They generalize the known results for optical cat states~\cite{Zurek93, dodonov00, SpohnDurr, Zurek2003} to all pure and mixed states.

\section{Effects of a large QCS} The expressions in~\eqref{eq:QCSWch} show that a large value of the QCS corresponds to a large spread of the characteristic function and  to the existence of small scale structures in the Wigner function~\cite{gu90, Bievre19}. Indeed, ${\||\xi|\chi\|_2^2}/{\|\chi\|_2^2}$  is the mean of $|\xi|^2$ with respect to the probability density ${|\chi(\xi)|^2}/{\|\chi\|_2^2}$. Hence, a large value of the QCS corresponds to a characteristic function with a wide spread in at least some directions in the $\xi$-plane, a manifestation of the well known link between the characteristic function and the coherences~\cite{leonhardt}:
\begin{equation*}
\chi(-\frac{\mu\sin\theta}{\sqrt2}, \frac{\mu\cos\theta}{\sqrt2})= \Tr \exp(i\mu X_\theta) 
=\hspace{-3pt}\int\hspace{-2pt} \rho(p_\theta, p_\theta+\mu)\rd p_\theta.
\label{eq:charfunctcoher}
\end{equation*}
On the other hand, a large QCS implies the gradient of $W$ is large, which means the graph of $W$ must have steep slopes, at least in some places of the phase plane, a signature either of oscillations or of sharp peaks~\cite{leje11, Bievre19}. For Gaussian states, this phenomenon manifests itself in that the variance of the probability distribution of one of the quadratures is of order $\Ccal^{-2}$ (more details in Appendix~\ref{gaussian}). A faithful reconstruction of the Wigner function through quantum tomography therefore requires great accuracy when $\Ccal\gg 1$. States with a large QCS are therefore hard to observe.  That it is generally difficult to measure  optical cat states and analogous states in other systems, when their components have a ``macroscopic'' separation, was proven in~\cite{segisa14}. We have here established the same result for all mixed or pure states of a bosonic quantum field with a large QCS.

\begin{figure}[]
		\hspace*{-250pt}\includegraphics[trim=6cm 13cm 5cm 13cm, clip, width=0.5\columnwidth]{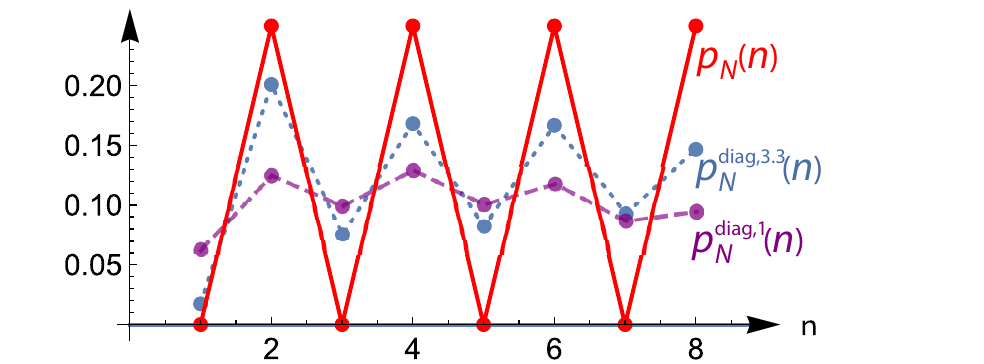}
	\hspace{5pt}
		\includegraphics[trim=6cm 12.6cm 5cm 12cm, clip,width=0.44\columnwidth]{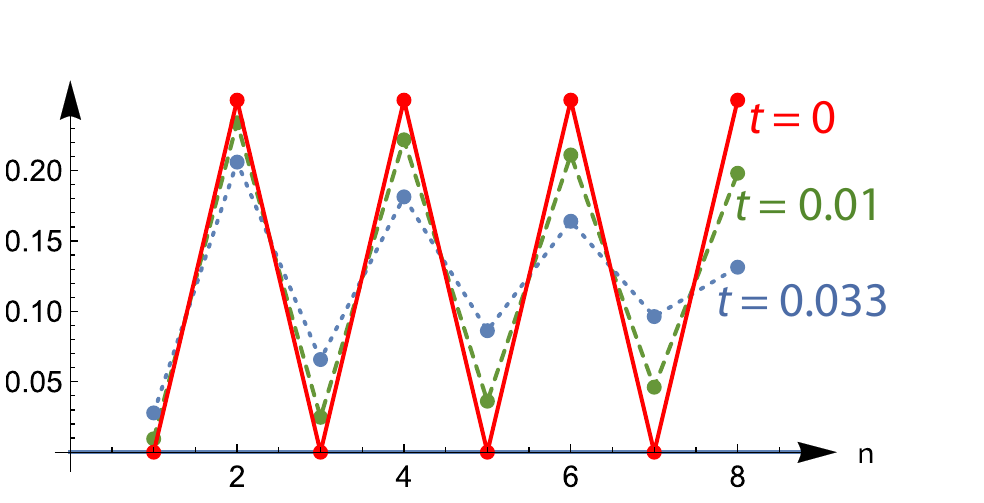}
	\caption{ Full red line: values of $p_N(n)=1/M$ for the even state $\rho_M$ with $M=4$ at $t=0$. Left panel: values of $p_N^{\textrm{diag},3.3}$ (dotted blue line) and $p_N^{\textrm{diag},1}$ (dashed purple line) as defined in~\eqref{eq:diagcoh}, both at $t=0$. Right panel: values of $p_N(n)$ at $t=0.01$ (dashed green line) and $t=0.033=\tau_\Ccal$ (dotted blue line). \label{fig:evencoherence}}
\end{figure}

To see how a large coherence scale can lead to strong interference effects, we consider the states $\rho_M$ (Fig.~\ref{fig:QCS}) and choose $A=X$ and $B=N=a^\dagger a$ and write
$$
p_N(n)= p_N^{\textrm{diag},\ell}(n) + \int_{|x'-x|\geq \ell} \langle x'|n\rangle \langle n|x\rangle\rho(x,x') \rd x\rd x',
$$
in analogy with~\eqref{eq:interference}. Here
\begin{equation}\label{eq:diagcoh}
p_N^{\textrm{diag}, \ell}(n)=\int_{|x'-x|\leq \ell} \langle x'|n\rangle \langle n|x\rangle\rho(x,x') \rd x\rd x'.
\end{equation}
Contrary to when $A$ has a discrete spectrum, as in~\eqref{eq:interference}, one cannot   sharply isolate the diagonal part of the state. Nevertheless, as the left panel of Fig.~\ref{fig:evencoherence} illustrates,
 it is the contribution of the coherences far from the diagonal that generate the sharp oscillations or fringes in $p_N(n)$.
In fact, it is clear (see Fig.~\ref{fig:evencoherence}) that the term $p_N^{\textrm{diag}, \ell}(n)$ shows a mildly oscillating behaviour for $\ell=1$, which is, as $\ell$ grows,  enhanced  by the interference terms to yield $p_N(2k)=1/M$ (constructive interference), $p_N(2k+1)=0$ (destructive interference). 
That the dynamical loss of large scale coherences  leads to a sharp decrease of this interference effect is illustrated in the right panel of Fig.~\ref{fig:evencoherence}: at the QCS half life $\tau_\Ccal=0.033$ of the state, the interferences are already considerably suppressed.


\section{Quadrature coherence scale and optical (non)classicality}  
Let $ \textrm{C}_{\textrm{cl}}$ 
be the set of 
optical classical states, \emph{i.e.} all mixtures of coherent states~\cite{tigl65}.  A number of witnesses, measures and monotones of optical nonclassicality have been designed~\cite{tigl65, hi85, balu86, hi87, hi89,  le91, agta92, le95, luba95, domamawu00, mamasc02, rivo02, kezy04, ascari05, se06, zapabe07, vosp14, ryspagal15, spvo15, kistpl16, al17, na17,  ryspvo17, yabithnaguki18, kwtakovoje19, Bievre19} 
to identify non optical classical states and to quantify their degree of non optical classicality. 
  Those quantities are often hard to compute, to measure, or to give a clear physical meaning. It is in particular not evident how they relate to standard manifestations of specifically quantum behaviour such as coherence and interference,  nor how they evolve when 
  coupled to a thermal bath. We show here a quantitative link between optical (non)classicality, the presence of coherences, and (fast) decoherence. Our analysis is based on the optical nonclassicality distance $d(\rho,\textrm{C}_{\textrm{cl}})$ defined in~\cite{Bievre19}  using  a quantity denoted $\OS(\rho)$ which measures the sensitivity of the state to operator ordering. 
One of its expression is 
$\OS(\rho)=\frac14\frac{\parallel \nabla W\parallel_2^2}{\parallel W\parallel_2^2}$. In view of~\eqref{eq:QCSWch}, this means
$\Ccal^2(\rho)=\OS(\rho)$. In other words, the QCS provides a new physical interpretation of the ordering sensitivity in terms of quadrature coherences, and the associated physical phenomena described above. In view of  the bound
\begin{equation}\label{eq:OSbound}
{\Ccal(\rho)}-1\leq d(\rho, \textrm{C}_{\textrm{cl}})\leq {\Ccal(\rho)}
\end{equation}
proven in~\cite{Bievre19}, $\Ccal(\rho)$ is a good estimate of the distance between $\rho$ and the optical classical states when $\Ccal(\rho)\gg1$. Hence the states far from the optical classical states are those with quadrature coherences far from the diagonal.
 In view of what precedes, they are the most fragile to decoherence. 
 Conversely, when $\rho\in\textrm{C}_{\textrm{cl}}$, $d(\rho,\textrm{C}_{\textrm{cl}})=0$ and it follows from~\eqref{eq:OSbound} that $\Ccal(\rho)\leq 1$: optical classical states are quadrature quasi-incoherent. Finally, the smaller the QCS of $\rho$, the closer it is to the optical classical states. This link between coherence and optical nonclassicality is specific to the quadrature coherences. It is for example not present in the $a^\dagger a$-coherence of the state.

\section{Conclusion} We 
introduced, for any bosonic state 
, its quadrature coherence scale (QCS), a measure of how far from the diagonal its quadrature coherences lie. We 
 established that the states with a large QCS are strongly optically nonclassical, hard to observe, and very sensitive to environmental decoherence. 
 These results generalize the known fast decoherence of ``large'' optical cat states~\cite{Zurek93, dodonov00, SpohnDurr, Zurek2003, hara13}  to all pure or mixed states with a large QCS.

One may thus legitimately argue that the QCS provides a measure of ``quantum macroscopicity''. Indeed, when the QCS is large, the state is ``strongly nonclassical'' in the sense that it is far from the optical classical states and its far off-diagonal coherences can be understood as a form of ``macroscopicity''. Also, when the QCS is small, the states are close to the optical classical states and in this sense have a low degree of ``quanticity''.  Our results thus strongly support a suggestion in~\cite{yabithnaguki18}, were it is surmised that there may be a link between optical nonclassicality and macroscopic quantum effects.

\acknowledgments
This work was supported in part by the Agence Nationale de la Recherche under grant ANR-11-LABX-0007-01 (Labex CEMPI) and by the Nord-Pas de Calais Regional Council and the European Regional Development Fund through the Contrat de Projets \'Etat-R\'egion (CPER). SDB thanks Prof. H. Spohn and J.C. Garreau for illuminating discussions on the subject matter of the paper. SDB thanks the CRM, where this work was initiated, for its hospitality in October-November 2018.

\appendix
%
%

\section{Quadrature quasi-incoherence}
\label{incoherent}

We defined in the Letter, for a state $\rho$ and observable $A$, the $A$-coherence scale (squared) $\Ccal_A^2(\rho)=\frac1{2\Pcal}\tr[\rho,A][A,\rho]$. A state is $A$-incoherent if $\Ccal_A(\rho)=0$. Choosing $A=X$, we now show there do not exist states that are $X$-incoherent. Indeed, 
\begin{equation}
\Ccal^2_X(\rho)=\frac1{\Pcal}\int\int (x-x')^2|\rho(x,x')|^2\rd x\rd x',
\end{equation}
so $\Ccal_X(\rho)=0$ implies $\rho(x,x')=\sigma(x)\delta(x-x')$. Hence, in the $X$-representation, $\rho$ is a multiplication operator: $\langle x|\rho|\psi\rangle=\sigma(x)\mean{x|\psi}$. Such an operator cannot have an orthonormal basis of eigenfunctions with corresponding eigenvalues $0\leq p_i\leq1$, $\sum_{i=1}^\infty p_i=1$. So $\Ccal_X(\rho)>0\,\,\,\forall\rho$.
It is clear, on the other hand, that $\Ccal_X(\rho)$ can be arbitrarily small. Note that, for $|x-x'|\gg \mathcal C_X(\rho)$, $\rho(x,x')$ must be small. In other words, states with a small $\Ccal_X(\rho)$ have appreciable $X$-coherences along the diagonal only. The same is true for any of the quadratures $X_\theta$ or $P_\theta$.

\section{Effective differential equations for $\Pcal(t)$ and $\Ccal(t)$}
\label{DiffEq}

In our analysis we consider the field is weakly coupled to a thermal bath through the standard master equation in Lindblad form~\cite{scla67, Joos1985,  alicki87, dodonov00, hara13}
\eq{\frac{d}{dt}\rho(t)=-i\omega [a^\dag a,\rho(t)]+\frac{1}{2}\gamma\left\{[a\rho(t),a^\dag]+[a,\rho(t)a^\dag]\right\}+\frac{1}{2}\delta\left\{[a^\dag\rho(t),a]+[a^\dag,\rho(t)a]\right\}\label{lindblad}}
where $\gamma>\delta\geq0$.
This model  is exactly solvable  in the Heisenberg picture. Indeed, one can show (see \cite{alicki87}) that the Weyl operator $D(\xi)=\e^{\xi a^\dag- \xi^* a}$ evolves in time as 
$	D(\xi;t)=	\e^{-\frac{1}{2}(2\bar n_\infty+1)|\xi|^2(1-\e^{-t/t_R})}
D(\xi_t)$ with $\xi_t=\e^{i\omega t-\frac{1}{2}t/t_R}\xi$
where we use the notation $D(\xi;0)\equiv~D(\xi)$ and defined $t_R=(\gamma-\delta)^{-1}$, the relaxation time and $\bar n_\infty=\delta t_R$,  the mean photon
number at infinity. 
Hence, $\chi(\xi;t)=\tr\rho(t)D(\xi)=\tr\rho D(\xi;t)$. It follows that
\eqarray{	\label{normchit}
	\|\chi(\xi;t)\|^2_2&=&\int |\chi(\xi;t)|^2\;\textrm{d}\xi
	=\int \e^{t/t_R-(2\bar n_\infty+1)(\e^{t/t_R}-1)|y|^2}|\chi(y)|^2\,\textrm{d}y,\\
	\||\xi|\chi(\xi;t)\|^2_2&=&\int |\xi|^2|\chi(\xi;t)|^2\;\textrm{d}\xi
	=\int \e^{2t/t_R-(2\bar n_\infty)(\e^{t/t_R}-1)|y|^2}\,|y|^2\,|\chi(y)|^2\,\textrm{d}y,
	\label{normchit2}}
where $|\xi|^2=|\xi_1|^2+|\xi_2|^2$ and $\textrm{d}\xi=\textrm{d}\xi_1\textrm{d}\xi_2$ (and similarly for $y$). Hence, the evolution of the quadrature coherence scale (QCS) is given by
\eq{\mathcal{C}^2(\rho(t))=\frac{\||\xi|\chi(\xi;t)\|^2_2}{\|\chi(\xi;t)\|^2_2}\label{S0t}.}
Given the initial state with characteristic function $\chi(\xi)$, (\ref{S0t}) can be computed numerically using \eqref{normchit}-\eqref{normchit2}. This is the way the graph of Fig. 2 of the Letter is produced.

\begin{figure}[]
	\begin{tikzpicture}
	\node[anchor=south west,inner sep=0] (image) at (-30,0) {			\includegraphics[width=0.4\columnwidth]{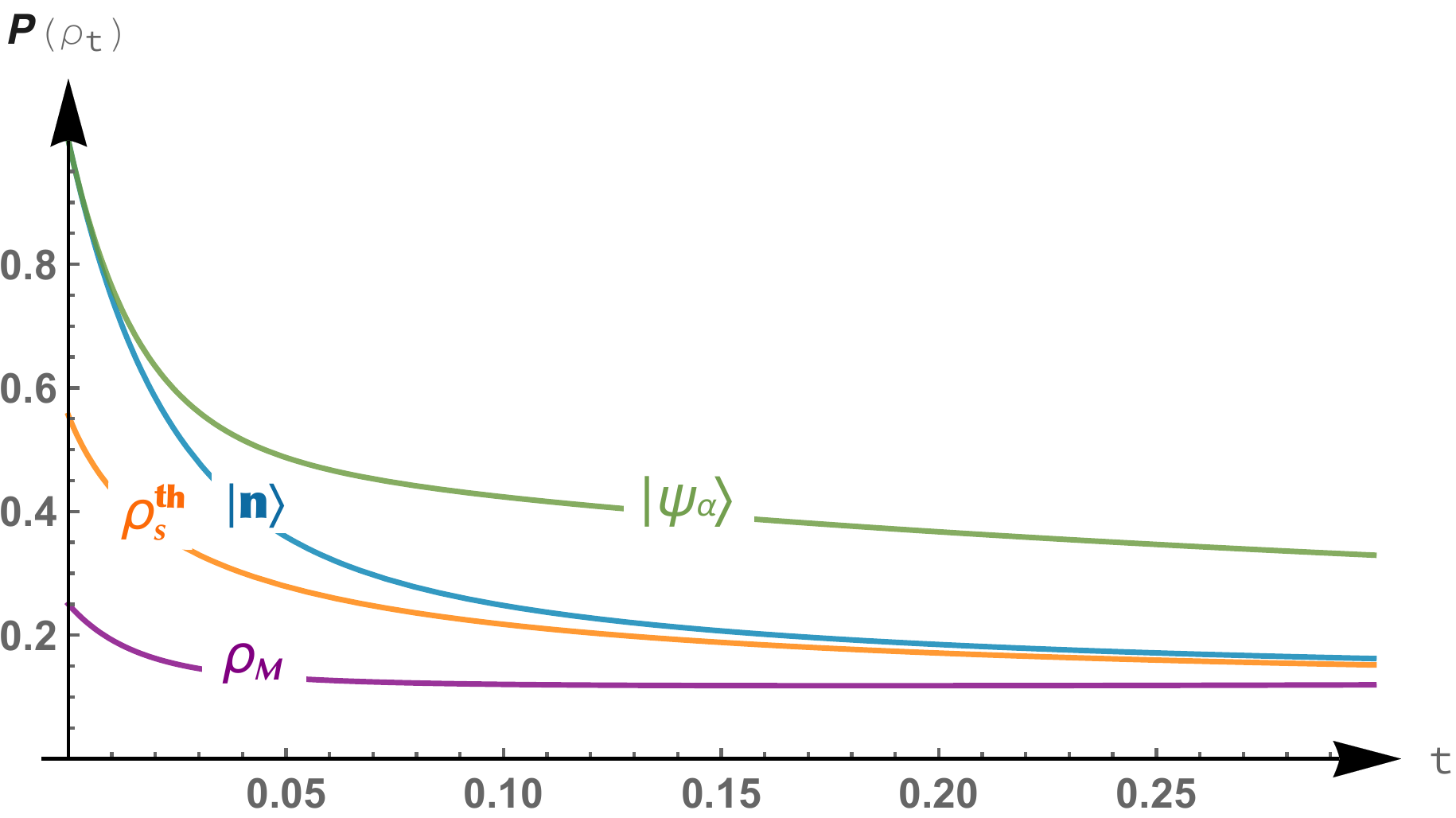}};
	\begin{scope}[x={(image.south east)},y={(image.north west)}]
	\node[anchor=south west,inner sep=0] (image) at (0.2,0.6) {\footnotesize
		\setlength{\tabcolsep}{0.2cm}	\begin{tabular}{ c||c|c|c|c| } 
		& $|n\rangle $ & $|\psi_\alpha\rangle $ & $\rho_M$& $\rho_s^{th}$\\ 
		\hline
		\hline
		$\tau_\mathcal{P}$  & 0.028& 0.045  & 0.067 & 0.050  \\ 
		\hline
		$\tau_{\mathcal{P},\textrm{appr.}}$& 0.016& 0.016 & 0.016& 0.052  \\ 
		\hline
		\end{tabular}\normalsize};
	\end{scope}
	\end{tikzpicture}
	\caption{\label{fig:evolutionPurity}Evolution of the purity $\mathcal{P}(\rho_t)$ of an initial Fock state $|n\rangle$ ($n=5$), a squeezed thermal state ($V=~1.8 \begin{psmallmatrix}
		e^{-2r}&0\\0&e^{2r}	\end{psmallmatrix},\,\, r=~\cosh^{-1}(19.8)/2\simeq 1.84$), an optical cat state $|\psi_\alpha\rangle\sim (|\alpha\rangle+|-\alpha\rangle)$ ($\alpha\simeq 2.24$) and a state $\rho_M$ with $M=4$. $\Ccal(0)=\sqrt{11}$ for all states shown. $\overline n_\infty = 1$. $t_R=1$. The table shows the exact value of the half-life $\tau_{\Pcal}$ of the purity and its approximation obtained by Eq.~\eqref{eq:halflifeP} for the first three columns and Eq.~\eqref{eq:halflifePurityGaussian} for the Gaussian state.}
\end{figure}

Similarly, the purity $\Pcal(t)=\Pcal(\rho(t))=\tr\rho(t)^2$ can be computed in terms of  the characteristic function as $\Pcal(t)=~\frac{1}{\pi}\|\chi(\xi;t)\|_2^2$ and its evolution is determined by 
\eq{\dot \Pcal(t)= \frac1{\tR}\left[1-(2\nbarinf+1)\Ccal^2(t)\right]\Pcal(t).\label{eq:evolutionPurity}} Figure \ref{fig:evolutionPurity} shows this evolution for three families of non-Gaussian states and one family of Gaussian states. 
With a linear approximation for small $t$,  the purity half time $\tau_{\Pcal}$ is  given by \eq{\tau_{\Pcal}\approx-\frac12\frac{\Pcal(0)}{\dot{\Pcal(0)}}=\frac12\frac{1}{(2\nbarinf+1)\Ccal^2(0)-1}\tR,\label{eq:halflifeP}}
provided  $\Ccal_0^2=\Ccal^2(0)>1$. It is inversely proportional to the square of the QCS and to the temperature of the bath (proportional to $\nbarinf$). This reduces to the known result for pure states~\cite{Zurek93, dodonov00, Zurek2003}  for which $\Ccal^2(0)$ coincides with the total noise of the initial state. As shown in the table of Fig.\ref{fig:evolutionPurity}, this approximation is not very accurate but gives the right order of magnitude. For Gaussian states, a better approximation can be obtained (see Eq.~(\ref{eq:halflifePurityGaussian})).

The analytical expression of $\Ccal(t)=\Ccal(\rho(t))$ obtained in \eqref{normchit}-\eqref{S0t} is not very informative. It does in particular not provide a simple expression for the time scale for coherence loss in term of the parameters of the model and of the initial condition. To remedy this situation, we 
establish the differential equation for $\mathcal{C} (t)$. First, we define  moments of $\xi_t$ as
\eq{\langle\langle\xi^{2k}\rangle\rangle_t=\int |\xi|^{2k} \frac{|\chi(\xi;t)|^2}{\|\chi(\xi;t)\|_2^2} \rd x\qquad\text{and}\qquad\kappa(t)=\kappa(\rho(t))=\left(\frac{\langle\langle\xi^4\rangle\rangle_t}{\langle\langle\xi^2\rangle\rangle_t^2}-1\right).\label{kappa}}
Note that $\langle\langle\xi^{0}\rangle\rangle_t=\|\chi(\xi;t)\|_2^2$ and $\langle\langle\xi^{2}\rangle\rangle_t=\mathcal{C}^2(t)$.
The derivative of  $\mathcal{C} (t)$ is then given by
\eqarray{\frac{d}{dt}\mathcal{C}^2(t)&=&\frac{1}{t_R}\mathcal{C}^2(t)+\frac{2\bar n_\infty+1}{t_R}(\mathcal{C}^4(t) -\langle\langle\xi_t^4\rangle\rangle )=\frac{1}{t_R}\Big(1-\kappa(t)(2\bar n_\infty+1)\mathcal{C}^2(t)\Big)\mathcal{C}^2(t)\nonumber\\
	\Leftrightarrow \quad\frac{d}{dt}\mathcal{C}(t)&=&\frac{1}{2t_R}\Big(1-\kappa(t)(2\bar n_\infty+1)\mathcal{C}^2(t)\Big)\mathcal{C}(t)\label{deriveeQCS}.}
This is Eq. (11) of the Letter.

\section{Gaussian states}
\label{gaussian}
\noindent\textbf{Definition.}
We consider a  Gaussian state $\rho_\mathrm{G}$ (centered on $0$). It is described by a covariance matrix
\eq{V=2\begin{pmatrix}
		\sigma_x^2&\sigma_{xp}\\\sigma_{xp}&\sigma_p^2
	\end{pmatrix}
	=\begin{pmatrix}
		2\tr\rho X^2&\tr\rho(XP+PX)\\\tr\rho(XP+PX)&2\tr\rho P^2
	\end{pmatrix}.}
Note that $\sigma_x^2=(\Delta X)^2$ and $\sigma_p^2=(\Delta P)^2$.
For the state to be physical we need $\det V\geq 1$ ($\hbar=1$).  The characteristic function of the Gaussian state is  given by \cite{weedbrook}
\eq{ \chi_\textrm{G}(\xi)=\e^{-\frac{1}{2}\xi^T\Omega V \Omega^T \xi}=\exp\left\{-(\sigma_p^2\xi_1^2+\sigma_x^2\xi_2^2+2\sigma_{xp}\xi_1\xi_2)\right\},\qquad\Omega=\begin{pmatrix}
		0&1\\-1&0
	\end{pmatrix}
	\label{characGaussien}
}
and its purity is $\mathcal{P}_\mathrm{G}=\tr(\rho^2_\mathrm{G})=\frac{1}{\pi}\|\chi_\mathrm{G}(\xi)\|^2_2=\frac{1}{\sqrt{\det V}}$.

\medskip
\noindent  \textbf{QCS of a Gaussian state.} The QCS of any Gaussian state can be computed with Eqs.~(\ref{normchit})-(\ref{S0t}): 
\eq{\mathcal{C}^2_\mathrm{G}=\mathcal{C}^2(\rho_\mathrm{G})
	=\frac{\sigma_x^2+\sigma_p^2}{\det V}=(\sigma_x^2+\sigma_p^2)\mathcal{P}^2=\frac12\Tr V^{-1}.\label{S0Gaussien}}
Since $\Ccal(\rho)$ is invariant under phase-space translations, this expression is valid  for all Gaussian states not necessarily centered at the origin.   	$\mathcal{C}_\mathrm{G}$ is also invariant under rotation.

One can diagonalize $V$ with a rotation of
angle $\theta_*$ so that $\sigma_{x_{\theta_*}p_{\theta_*}}= 0$. There exist two such angles, but we chose it such that $\sigma_{x_{\theta_*}}^2\leq\sigma_{p_{\theta_*}}^2$.
We have from Eq.~(\ref{S0Gaussien}) that
$	\QCStwoG=\frac14\left(\frac1{\sigma_{\xthetastar}^2}+\frac1{\sigma_{\pthetastar}^2}\right).$
Hence,
\eq{	\frac1{2\sigma_{\pthetastar}^2}=\frac12\min\left\{\frac1{\sigma_{\xthetastar}^2},\frac1{\sigma_{\pthetastar}^2}\right\}\leq \QCStwoG\leq\frac12\max\left\{\frac1{\sigma_{\xthetastar}^2},\frac1{\sigma_{\pthetastar}^2}\right\}= \frac1{2\sigma_{\xthetastar}^2}.\label{minmax}}
In particular, if $\QCStwo_{\mathrm G}\gg1$, then $\sigma_{\xthetastar}$ is small so that the density of the quadrature $X_{\theta_*}$ is sharply peaked. This shows that Gaussian states with a large $\Ccal_{\mathrm{G}}$ are hard to measure.
By the uncertainty principle $\sigma_{\xthetastar}^2\sigma_{\pthetastar}^2\geq \frac14$~\cite{Heisenberg, Kennard}; the density of $P_{\theta_*}$ must then be very broad. In fact,
\eq{  	\frac1{4\sigma_{\pthetastar}^2}
	\oset{\text{Uncertainty principle}}{\leq}
	\sigma_{\xthetastar}^2	\oset{\text{From Eq.~(\ref{minmax})}}{\leq} \frac1{2\QCStwo_{\mathrm G}},}
from which we deduce in particular that $\frac{\QCStwo_{\mathrm G}}{2}\leq  \sigma_{\pthetastar}^2$. Thus, if $\QCStwo_{\mathrm G}\geq 1$, then 
$	\sigma_{\xthetastar}^2\leq \frac1{2\QCStwo_{\mathrm G}}\leq\frac{\QCStwo_{\mathrm G}}{2}\leq  \sigma_{\pthetastar}^2$.
Then $\Ccal_{\textrm{G}}\simeq (2\sigma_{\xthetastar})^{-1}$. 
The link with the support of the coherences can be made explicitly as follows. The characteristic function is linked to the coherences through~\cite{leonhardt}:
\begin{equation}
\chi(-\frac1{\sqrt2}\mu\sin\theta, \frac1{\sqrt2}\mu\cos\theta)= \Tr \exp(i\mu X_\theta) =\int \rho(p_\theta, p_\theta+\mu)\rd p_\theta.
\label{eq:charfunctcoher}
\end{equation}
Since the left-hand side of~\eqref{eq:charfunctcoher} behaves as $\exp(-\frac12\mu^2\sigma_{\xthetastar}^2)$, its spread is wide, of order $\sigma_{\xthetastar}^{-1}$. This implies the $\rho(\pthetastar, {\pthetastar}+\mu)$ coherences in the right-hand side have a support in $\mu$ of the same order. The coherences $\rho({\xthetastar}, {\xthetastar}+\mu)$, on the other hand, live on the small scale $\sigma_{\pthetastar}^{-1}$.

We can also easily compute from \eqref{kappa}
\eq{\kappa_\mathrm{G}=\kappa(\rho_{\mathrm{G}})=2-\frac{\det V}{(\sigma_x^2+\sigma_p^2)^2}.\label{kappaG}}
Note that $1\leq\kappa_\mathrm{G}\leq2$ and in particular, $\kappa_\mathrm{G}=1$ for all thermal state while it tends to 2 when the squeezing becomes large.
Table \ref{exampleGaussian} lists the values of the QCS, the purity and $\kappa_\mathrm{G}$ for some specific examples of Gaussian states.

\begin{table}
	\begin{center}
		\renewcommand{\arraystretch}{1.9}
		\begin{tabular}{ c||c|c|c| } 
			&Coherent& Thermal & Squeezed thermal \\ 
			\cline{2-4} 
			& \multirow{2}{*}{$\ket{\alpha}=D(\alpha)\ket{0}$}&$\rho_{th}=(1-q)\sum_{n=0}^{\infty}q^n\ketbra{n}{n}$ & $\rho_{th,s}=S\rho_{th}S^\dag$  \\ 
			& &$\bar n=\Tr \rho_{th}a^\dag a=(q^{-1}-1)^{-1}$ &  $S=\e^{\frac12(z^*a^2-za^{\dag2})},\quad z=e^{i\phi r}$  \\ 
			\hline
			\hline
			$\mathcal{C}_\mathrm{G}$& 1& $\frac{1}{\sqrt{1+2\bar n}}$ & $\sqrt{\frac{\cosh(2r)}{1+2\bar n}}$   \\ 
			\hline
			$\mathcal{P}_\mathrm{G}$ & 1&$ \frac{1}{1+2\bar n}$ &$\frac{1}{1+2\bar n} $\\ 
			\hline
			$\kappa_\mathrm{G}$ & 1&1 &$2-\frac{1}{\cosh^2(2r)} $\\ 
			\hline
		\end{tabular}
	\end{center}
	\caption{$\mathcal{C}_\mathrm{G}$, $\mathcal{P}_\mathrm{G}$, and $\kappa_\mathrm{G}$ for three families of Gaussian states.\label{exampleGaussian}}
\end{table}

\medskip 
\noindent\textbf{Evolution of the QCS.}  Since a Gaussian state remains Gaussian during the time evolution, both $\mathcal{C}_\mathrm{G}(t)$ and $\kappa_\mathrm{G}(t)$ can be evaluated using the covariance matrix at time $t$ given by
\eq{V(t)=\e^{-t/t_R}\begin{pmatrix}
		\cos \omega t&\sin \omega t\\-\sin \omega t&\cos \omega t
	\end{pmatrix} V \begin{pmatrix}
		\cos \omega t&-\sin \omega t\\ \sin \omega t&\cos \omega t
	\end{pmatrix} +(2\bar n_\infty+1)\left(1-\e^{-t/t_R}\right)\mathds{1}\label{Vt}}
and Eqs.~(\ref{S0Gaussien})-(\ref{kappaG}).
The exact expression of the QCS obtained in this manner is however not easy to interpret. It turns out that for Gaussian states $\kappa_\mathrm{G}(t)$ evolves slowly, a fact that can be anticipated by the observation that $1\leq\kappa_\mathrm{G}(t)\leq2$ for all $t$ and that can also be confirmed numerically.
Now, assuming $\kappa_\mathrm{G}(t)\simeq\kappa_\mathrm{G}(0)$ the solution of Eq.~(\ref{deriveeQCS}) is easily computed:
\eq{\mathcal{C}^2_\mathrm{G}(t)=\frac{\mathcal{C}_0^2}{\e^{-t/t_R}+\mathcal{C}_0^2\kappa_0(2\bar n_\infty+1)(1-\e^{-t/t_R})}.\label{approx}} where $\mathcal{C}^2_\mathrm{G}(0)\equiv\mathcal{C}_0$ and $\kappa_\mathrm{G}(0)\equiv\kappa_0$.
From it, we deduce the QCS half life, that is the the time $\tau_{\Ccal,\mathrm{G}}$ when $\mathcal{C}_\mathrm{G}(\tau_{\Ccal,\mathrm{G}})=~\mathcal{C}_0/2$ (provided  $\Ccal_0^2\kappa_0(2\nbarinf+1)>4$) and the time $\tau_{1,\mathrm{G}}$ at which the state becomes quasi-incoherent, that is when $\mathcal{C}(\tau_{1,\mathrm{G}})=1$ (provided  $\Ccal_0^2>1$):
\eq{\tau_{\Ccal,\mathrm{G}}=\ln\left(1+\frac{3}{\mathcal{C}_0^2\kappa_0(2\bar n_\infty+1)-4}\right)t_R,\qquad\tau_{1,\mathrm{G}}=\ln\left(1+\frac{1-1/\mathcal{C}_0^2}{\kappa_0(2\bar n_\infty+1)-1}\right)t_R.}
If $\bar n_\infty$ is fixed and $\mathcal{C}_0$ is large (high QCS), then
\eq{\tau_{\Ccal,\mathrm{G}}\approx\frac{3}{\mathcal{C}_0^2\kappa_0(2\bar n_\infty+1)-4}t_R\quad \text{and} \quad\tau_{1,\mathrm{G}}\approx\left(\ln\left(\frac{\kappa_0(2\bar n_\infty+1)}{\kappa_0(2\bar n_\infty+1)-1}\right)-\frac{1}{\mathcal{C}_0^2\kappa_0(2\bar n_\infty+1)}\right)t_R.} Both times are inversely proportional to the square of the QCS, $\kappa_0$ and to the temperature of the bath (proportional to $\bar n_\infty$).

\begin{figure}[]
	\begin{tikzpicture}
	\node[anchor=south west,inner sep=0] (image) at (-70,0) {		\includegraphics[width=0.4\columnwidth]{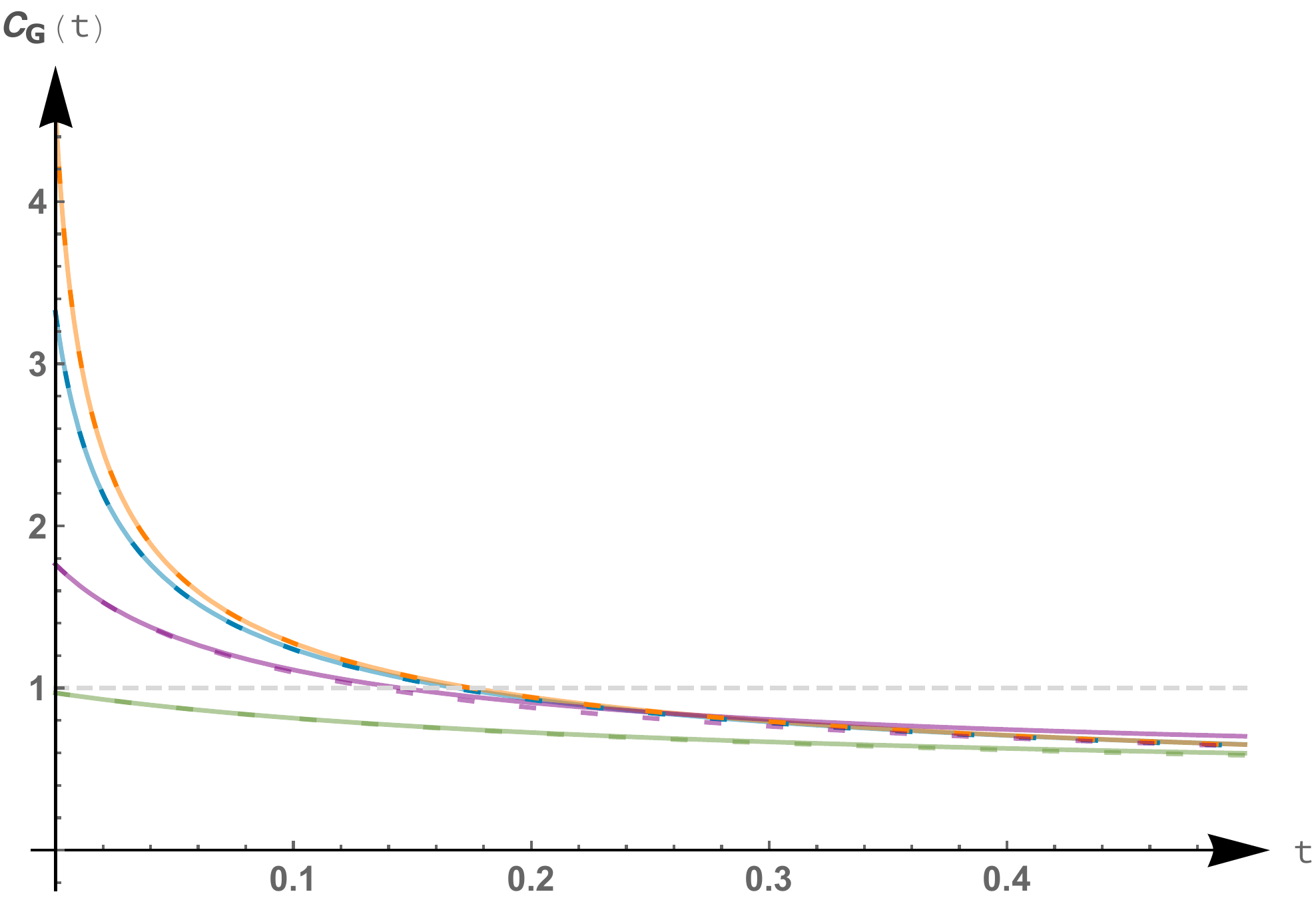}};
	\begin{scope}[x={(image.south east)},y={(image.north west)}]
	\node[anchor=south west,inner sep=0] (image) at (0.3,0.6) {\footnotesize
		\setlength{\tabcolsep}{0.2cm}	\begin{tabular}{ c|c|c } 
		$\beta$& $r$&$\mathcal{C}_\mathrm{G}(0)$\\ 
		\hline\hline
		0.9&2.2&4.58\\
		0.9&1.8&3.32\\
		0.5&0.9&1.76\\
		2&1&0.96\\
		\hline
		\end{tabular}\normalsize};
	\end{scope}
	\end{tikzpicture}
	\caption{\label{fig:comparisonGaussien} Evolution of the QCS of squeezed thermal (Gaussian) states with covariance matrix $V=\beta\begin{psmallmatrix}
		\e^{-2r}&0\\0&\e^{2r}
		\end{psmallmatrix}$. Values of $\beta$ and $r$ are given in the table in decreasing order of initial QCS. Plain lines show the exact evolution (obtained from \eqref{S0Gaussien}, \eqref{kappaG} and \eqref{Vt}) and dashed lines its approximation \eqref{approx}. }
\end{figure}

Figure \ref{fig:comparisonGaussien} shows the evolution of $\mathcal{C}_\mathrm{G}(t)$ for different Gaussian states (we only consider squeezed thermal states which have diagonal covariance matrix as $\mathcal{C}_\mathrm{G}$ is invariant under rotations). Plain lines show the exact evolution while the approximation obtained with Eq.~(\ref{approx}) is shown by the dashed lines. As we can see Eq.~(\ref{approx}) provides a very good estimation of $\mathcal{C}_\mathrm{G}(t)$, especially for short times. 

\medskip
\noindent\textbf{Evolution of the purity.}  Knowing the evolution of the QCS for a Gaussian state, it is possible to compute the evolution of its purity $\Pcal_\textrm{G}(t)$ by solving Eq.~(\ref{eq:evolutionPurity}) where one replaces $\Ccal$ by $\Ccal_{\mathrm{G}}$ given in Eq.~(\ref{approx}). The solution is given by \eq{\mathcal{P}_\textrm{G}(t)=\Pcal(0)\e^{t/\tR}\left(1+\Ccal_0^2\kappa_0(2\nbarinf+1)(\e^{t/\tR}-1)\right)^{-1/\kappa_0}.}
The half life time $\tau_{\Pcal,\textrm{G}}$ can then be obtained by solving 
\eq{1+z=2^{\kappa_0}\left(1+\frac{z}{L}\right)^{\kappa_0}\qquad\text{where}\qquad z=L(\e^{\tau_{\Pcal,\textrm{G}}/\tR}-1),\qquad L=\Ccal_0^2\kappa_0(2\nbarinf+1).}
For high QCS, $z/L$ is small. Assuming $\left(1+\frac{z}{L}\right)^{\kappa_0}\approx1+\kappa_0\frac{z}{L}+\frac{\kappa_0(\kappa_0-1)}{2}\left(\frac{z}{L}\right)^2$ we find
\eq{\tau_{\mathcal P, \mathrm{G}}\approx\ln\left(1+\frac{2^{\kappa_0}-1}{\left((2\overline n_\infty +1)\Ccal_0^2-2^{\kappa_0}\right)\kappa_0}\right)\tR\approx\frac{2^{\kappa_0}-1}{\left((2\overline n_\infty +1)\Ccal_0^2-2^{\kappa_0}\right)\kappa_0}\tR\label{eq:halflifePurityGaussian}}
provided that $(2\nbarinf+1)\Ccal_0^2>2^{\kappa_0}$.
One observes furthermore that for $\Ccal_0^2\gg1$, $\tau_{\Pcal, \mathrm G}\simeq \frac{3}{2^{\kappa_0}-1}\tau_{\Ccal, \mathrm{G}}$ which then implies $\tau_{\Pcal, \mathrm{G}}\leq \tau_{\Ccal, \mathrm{G}}\leq 3\tau_{\Pcal, \mathrm{G}}$: the purity loss is always slightly faster than the coherence scale loss.

\medskip
\noindent\textbf{Quantum Fisher information.}  The Quantum Fisher Information (QFI) $\mathcal F(\rho, A)$ of the state $\rho$ for the observable $A$  is defined as~\cite{brca94}
$
\mathcal F(\rho, A)=4\partial_x^2 D_B^2(\rho, \exp(-ixA)\rho\exp(ixA)_{|x=0},
$
where $D_B^2(\rho,\sigma)=2(1-F(\rho, \sigma))$ is the Bures distance and $F(\rho,\sigma)=\Tr \sqrt{\sqrt{\rho}\sigma\sqrt{\rho}}$ the fidelity between $\rho$ and $\sigma$. 
It was proven in \cite{yabithnaguki18} that for Gaussian states 
\eq{\frac14\Big(\mathcal F(\rho, X)+\mathcal F(\rho, P)\Big)=\frac12\tr V^{-1}.}
Following Eq.~(\ref{S0Gaussien}), this is equal to the QCS squared of a Gaussian state i.e. $\Ccal_{\mathrm{G}}^2$.

\section{Computation of the QCS for some pure  states}
\label{purestates}

\medskip
\noindent\textbf{Fock State $\ket{n}$.}
As a pure state, its QCS is given by
$ \mathcal{C}_n=\mathcal{C}(\ket{n}\bra{n})=\sqrt{(\Delta X)^2+(\Delta P)^2}=\sqrt{2n+1}.$ 
Its characteristic function is 
$\chi_n(\xi)=\e^{-\frac{|\xi|^2}{2}}L_n(|\xi|^2)$ where $L_n$ are the Laguerre polynomials. 
Using Eq.~(\ref{kappa}), one can compute $\kappa_n$ of a Fock state:
\eq{\kappa_n=\frac{2n^2+2n+1}{4n^2+4n+1}\qquad\text{with}\qquad \frac{1}{2}\leq\kappa_n\leq1.}
The integrals involved in the calculations can be computed analytically by using radial coordinates and Eq.~7.414(12) on p.~809 in \cite{gradshteyn2007}.

\medskip
\noindent\textbf{Cat state.}
The QCS of a cat state $\ket{\psi_\alpha}=\frac{1}{\sqrt{\mathcal{N}}}\left(\ket{\alpha}+\ket{-\alpha}\right)$ where $\ket{\alpha}$ is a coherent state and $\mathcal{N}=2(1+\e^{-2|\alpha|^2})$ is given by
$\mathcal{C}_\alpha=\mathcal{C}(\ketbra{\psi_\alpha}{\psi_\alpha})=\sqrt{(\Delta X)^2+(\Delta P)^2}=\sqrt{1+2|\alpha|^2\tanh|\alpha|^2}.$
Its characteristic function is
$$\chi_\alpha(\xi)=\frac{1}{\mathcal{N}}\left(\e^{-|\xi|^2/2}(\e^{\xi\bar{\alpha}-\bar \xi\alpha}+\e^{-(\xi\bar \alpha-\bar \xi \alpha)})+\e^{-|2\alpha+\xi|^2/2}+\e^{-|2\alpha-\xi|^2/2}\right).$$
Using Eq.~(\ref{kappa}), one can easily compute 
\eq{\kappa_\alpha=1+\frac{4|\alpha|^4}{(\cosh|\alpha|^2+2|\alpha|^2\sinh|\alpha|^2)^2}\qquad\text{with}\qquad1\leq\kappa_\alpha\leq2.} 


\bibliographystyle{apsrev4-1}

%

\end{document}